\documentclass[letter]{aa}

\newcommand{\se}[1]{Section\ \ref{sec:#1}}

\usepackage{graphicx}
\usepackage{subfigure}

\usepackage{txfonts}

\begin{document}

   \title{Steady state by recycling prevents premature collapse of protoplanetary atmospheres}

   \author{T. W. Moldenhauer
          \inst{1}
          \and
          R. Kuiper
          \inst{1}
          \and
          W. Kley
          \inst{1}
          \and
          C. W. Ormel
          \inst{2}
         }

   \institute{Institut für Astronomie und Astrophysik, Universität Tübingen,
              Auf der Morgenstelle 10, 72076 Tübingen, Germany
              \and
              Department of Astronomy, Tsinghua University,
              30 Shuangqing Rd, Haidian District, Beijing, China\\\\
              \email{tobias.moldenhauer@uni-tuebingen.de}
             }

   \date{Received 2020; accepted 2021}
   
\titlerunning{Steady state by atmospheric recycling}
\authorrunning{Moldenhauer et al.}
 
  \abstract
  {
   In recent years, space missions such as \textit{Kepler} and \textit{TESS} have discovered many close-in planets with significant atmospheres consisting of hydrogen and helium: mini-Neptunes.
   This indicates that these planets formed early in gas-rich disks while avoiding the runaway gas accretion that would otherwise have turned them into hot-Jupiters.
   A solution is to invoke a long Kelvin-Helmholtz contraction (or cooling) timescale, but it has also been suggested that thermodynamical cooling can be prevented by hydrodynamical planet atmosphere-disk recycling.
  }
   {We investigate the efficacy of the recycling hypothesis in preventing the collapse of the atmosphere, check for the existence of a steady state configuration, and determine the final atmospheric mass to core mass ratio.}  
   {We use three-dimensional radiation-hydrodynamic simulations to model the formation of planetary proto-atmospheres.
   Equations are solved in a local frame centered on the planet.}
   {Ignoring small oscillations that average to zero over time, the simulations converge to a steady state where the velocity field of the gas becomes constant in time.
   In a steady state, the energy loss by radiative cooling is fully compensated by the recycling of the low entropy gas in the planetary atmosphere with high entropy gas from the circumstellar disk.}
   {For close-in planets, recycling naturally halts the cooling of planetary proto-atmospheres, preventing them from contracting toward the runaway regime and collapsing into gas giants.}


\maketitle

\section{Introduction}
Close-in planets are a new exoplanet population that were discovered in abundance by the \textit{Kepler} mission and now also by the Transiting Exoplanet Survey Satellite (\textit{TESS}). 
They tend to be small, up to several Earth radii in size, and are located close to their host stars in systems with a fair degree of uniformity \citep{WeissEtal2018}. 
Above all, they are very common: One out of three solar-type stars may host these types of planets \citep{WinnFabrycky2015,ZhuEtal2018}. 
Important to this work, their atmospheres are inferred to harbor significant amounts of hydrogen and helium \citep{WuLithwick2013}, strongly suggesting that they formed early within the primordial disk, hence the name "mini-Neptune." 
The observed valley in the planet size histogram \citep{FultonEtal2017} further suggests that even those planets without an atmosphere were born in the disks, with the smallest and most irradiated of them having their atmospheres stripped over time \citep{JinMordasini2018}.

From a theoretical perspective, there is no consensus on how exactly the atmosphere was accreted. Because these planets orbit so close to their host star, they likely assembled in a very short time \citep{Lee_2014}. Even if they did not form \textit{in situ} but migrated to their present orbital radii, it means that their proto-atmospheres had copious time to contract on the Kelvin-Helmholtz timescale, increasing their atmosphere mass to eventually become self-gravitating and form hot-Jupiters \citep{BatyginEtal2016}.  The rapid thermal evolution follows from our expectation that these atmospheres are characterized by a low opacity due to the fact that grains settle quickly \citep{Ormel2014,Mordasini2014} and are not replenished by planetesimal infall \citep{MovshovitzEtal2010}.  But for mini-Neptunes (and super-Earths), which by far outnumber hot-Jupiters, this runaway gas accretion must have been avoided. Proposed solutions include a high opacity atmosphere \citep{Lee_2014}, late formation \citep{LeeChiang2015}, atmospheric recycling \citep{OrmelEtal2015i, Ali-DibEtal2020}, and gap opening \citep{Fung_2018, Ginzburg-ea-2019}.

In this letter, we report that planet atmosphere-disk interaction yields a steady state and results in the complete recycling of the atmosphere. The recycling hypothesis states that atmospheres are not isolated from their disks, which may be inferred (erroneously) from one-dimensional or two-dimensional modeling. Isothermal three-dimensional simulations by \citet{OrmelEtal2015i}, \citet{FungEtal2015}, and \citet{Bethune-Rafikov-2019} instead have shown that disk gas flows deep into the atmosphere and vice versa. However, the flow structure was seen to strongly depend on the thermodynamical properties of the gas, that is, its cooling rate \citep{KurokawaTanigawa2018,Popovas_2018}. Therefore, this problem is best addressed with radiation hydrodynamical (RHD) simulations \citep{Ayliffe-Bate2009, Cimerman_2017}. Here, we present the results of new three-dimensional high resolution RHD simulations of an Earth-mass core embedded in a disk and demonstrate that the atmosphere reaches an equilibrium.
A true steady state is retrieved, in which the atmosphere mass reaches an asymptotic value, because gas elements are transported back to the disk before they have the chance to cool. 

\section{Model setup}
\label{sec:model}
To study the physics of protoplanetary atmospheres, we 
followed \citet{Cimerman_2017} and ran RHD simulations on a three-dimensional grid in spherical coordinates in a local shearing frame centered on the planet.
We inserted a planetary core with a mass $M_c = 1 \, M_\mathrm{Earth}$ at a distance $a = 0.1 \, \mathrm{au}$ from the star in the protoplanetary disk and followed the formation of an atmosphere around the core.
Following \cite{Lee_2014}, assuming a minimum-mass extrasolar nebula \citep[MMEN,][]{Chiang_2013}, we adopted an ambient midplane density of $\rho_0 = 6\cdot10^{-6} \, \mathrm{g}\,\mathrm{ cm}^{-3}$ and an ambient temperature of $T_0 = 1000 \, \mathrm{K}$.
The initial temperature is constant, and the density is vertically Gaussian.
For the rocky core, we took a mean density of $\rho_c = 5 \, \mathrm{g} \, \mathrm{cm}^{-3}$.
In units of the thermal mass $M_{th} = c_s^3/(G\Omega_K)$ \citep{Rafikov_2006}, the planetary core mass corresponds to a dimensionless mass of $m = M_c/M_{th} = 0.18$. For the time unit, we used the inverse Keplerian orbital frequency of the planet with respect to the star, $\Omega_K^{-1} = \sqrt{a^3 / (G M_\star)}$.

In our numerical setup, we followed \cite{Cimerman_2017} and used the open source PLUTO code \citep{Mignone_2007} and the flux-limited diffusion solver from \cite{Kuiper_2009} and \cite{Kuiper_2020}.
Similar to \citep{OrmelEtal2015i} and \citet{Cimerman_2017}, we adopted a spherical grid centered on the planet, but contrary to \cite{Cimerman_2017}, we simulated the full upper hemisphere around the core, assuming symmetry with respect to the equatorial plane. Previously, we had used the additional symmetry of the local shear and restricted the computational domain to half of the upper hemisphere. However, in this case, gravitational smoothing was necessary to avoid numerical instabilities caused by the boundaries together with the radiative transfer.
By simulating the full upper hemisphere in our new setup, the uncorrected Newtonian potential for the gravity of the planetary core can be used throughout, which stabilizes the flow structure of the inner atmosphere.
Our new simulations therefore no longer feature these unphysical flow artifacts.

For a steady state to emerge, the simulations have to exceed the Kelvin-Helmholtz timescale. Hence, in order to keep the cooling timescale short and thus save on computation time, we used a low constant opacity of $\kappa = 10^{-4} \, \mathrm{cm}^2 \, \mathrm{g}^{-1}$.
At that opacity, the protoplanetary atmosphere is still optically thick.
For the spherical grid, we used 128 logarithmically spaced grid cells in the radial direction, 32 equidistant cells in the polar direction, and 128 cells in the azimuthal direction.
We tested for convergence by comparing the density profile against a simulation with half resolution.
In the lower resolution simulation, the density profile differs only slightly in its evolution and settles into the same steady state.
As a comparison, and to demonstrate the impact of recycling, we also ran a one-dimensional simulation with otherwise identical parameters and physics.

For the inner boundary at the planetary core, we used a closed boundary condition where all fluxes, including the radiative, vanish (i.e., no energy is exchanged with the planetary core).
At the outer boundary, the hydrodynamical quantities were kept constant at their initial values to account for a circumstellar disk that evolves on much longer timescales.
The coordinate system is oriented in such a way that the vertical, $\theta=0$, axis points in the direction of the disk's angular momentum vector.
In a local Cartesian system, this corresponds to the $z$-axis, where the $x$-axis points to the radial direction (away from the star) and the $y$-axis in the direction of the disk's motion.

In order to prevent initial shocks at the planetary core surface and allow for a smooth simulation start, we multiplied the planet's gravity by a switch-on factor,
$\alpha_\mathrm{inj}(t) = 1-\exp( -0.5 t^2/t_\mathrm{inj}^2)$,
where $t_\mathrm{inj} = 4 \, \Omega_K^{-1}$ is the injection time.
For later usage, we defined the Hill radius as $R_\mathrm{Hill} = a (M_c/(3M_\star))^{1/3}$.

\begin{figure}[htbp]
   \centering
   \includegraphics[width=0.98\hsize]{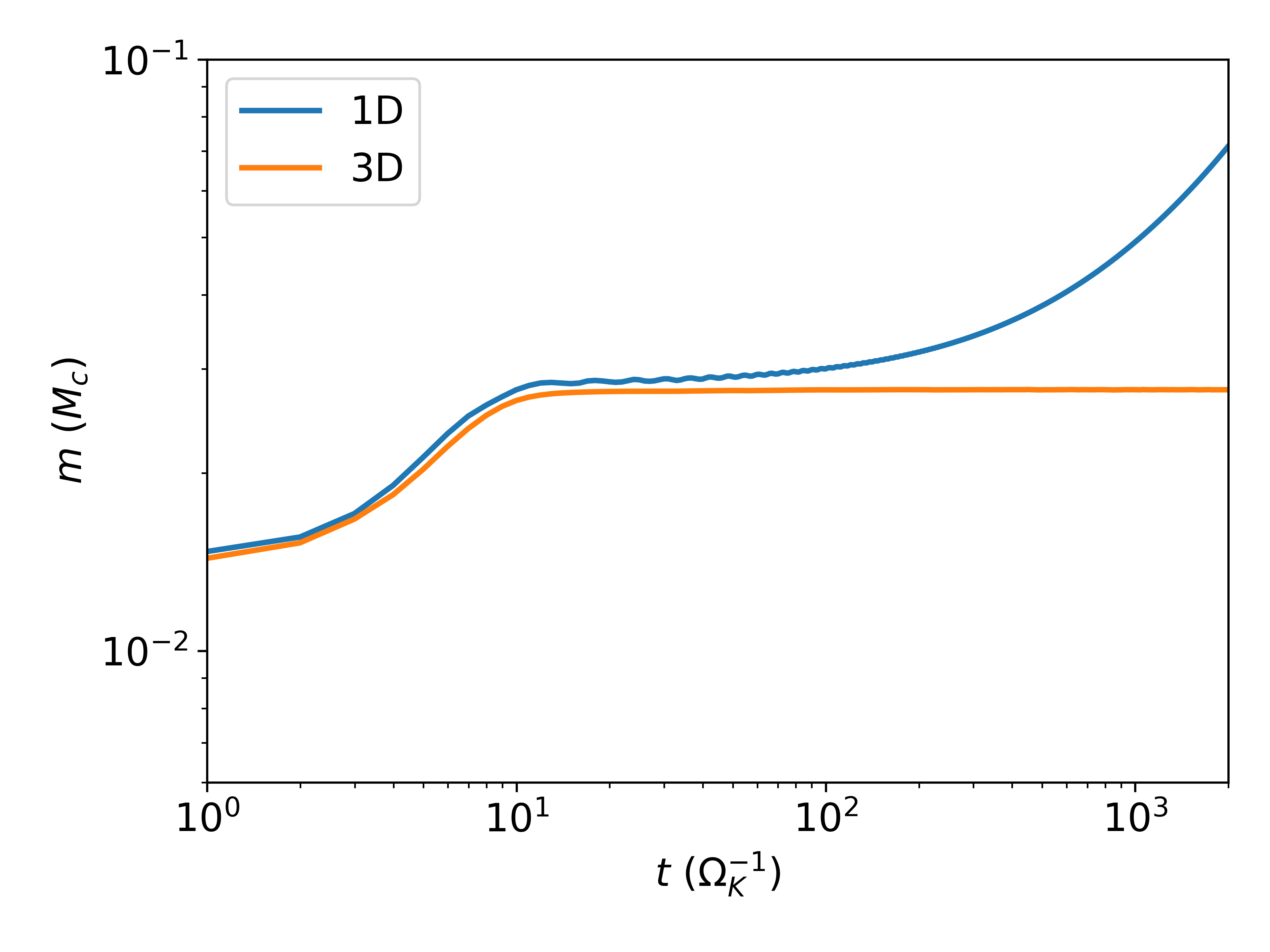}
   \caption{
      Mass evolution comparison inside the Hill sphere for the one-dimensional (blue) simulation and the three-dimensional (orange) simulation.
      The one-dimensional simulation starts with a slightly higher mass than the three-dimensional simulation because it does not account for the vertical stratification of the circumstellar disk.
      While the one-dimensional simulation continues to accrete, the three-dimensional simulation reaches a steady state, where the final mass, $\chi = M_\mathrm{final}/M_c = 2.76\,\%$, is typical for a mini-Neptune planet.
   }
   \label{fig:mass}%
\end{figure}

\begin{figure*}[htbp]
   \centering
   \includegraphics[width=0.49\hsize]{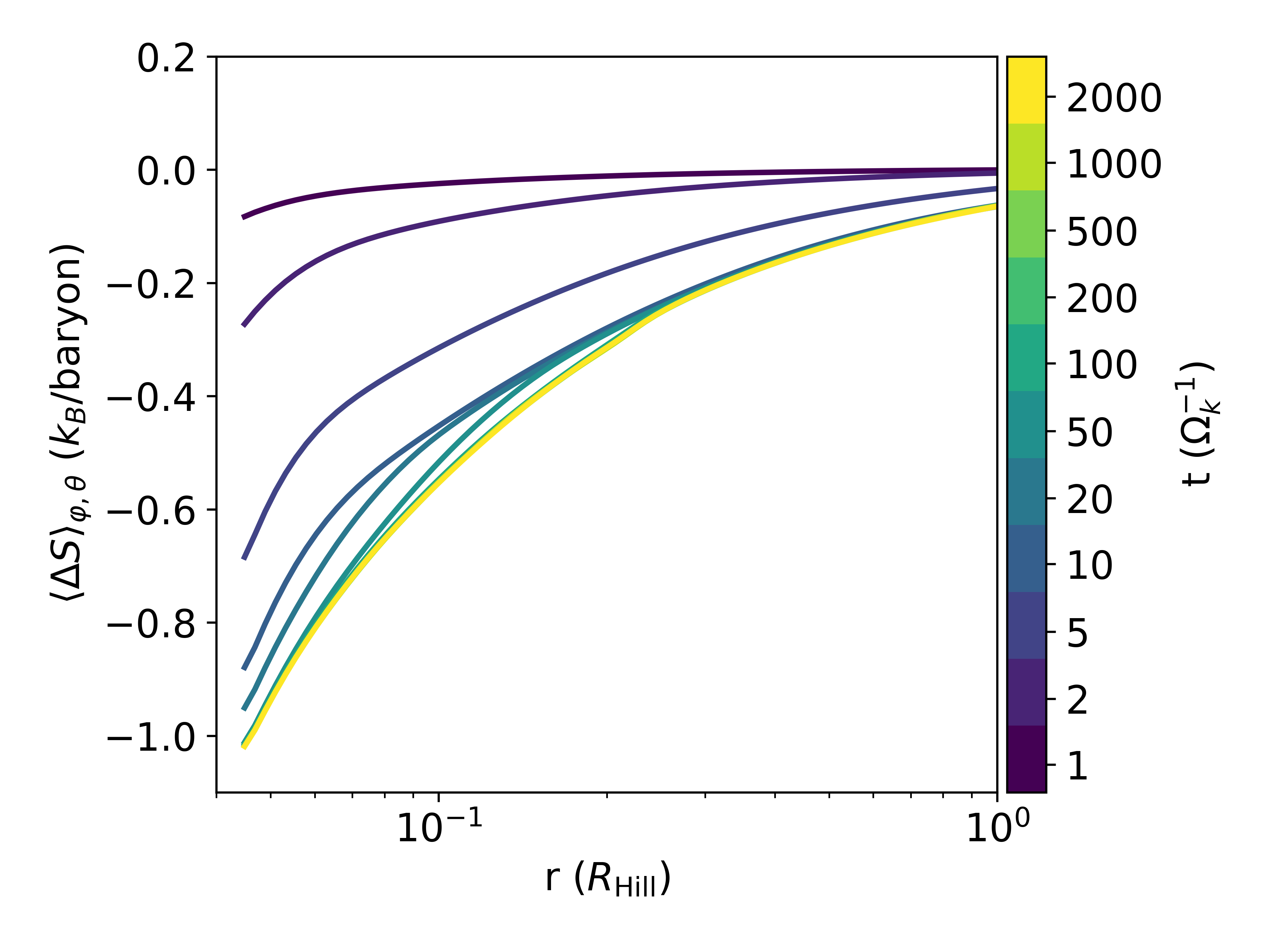}
   \includegraphics[width=0.49\hsize]{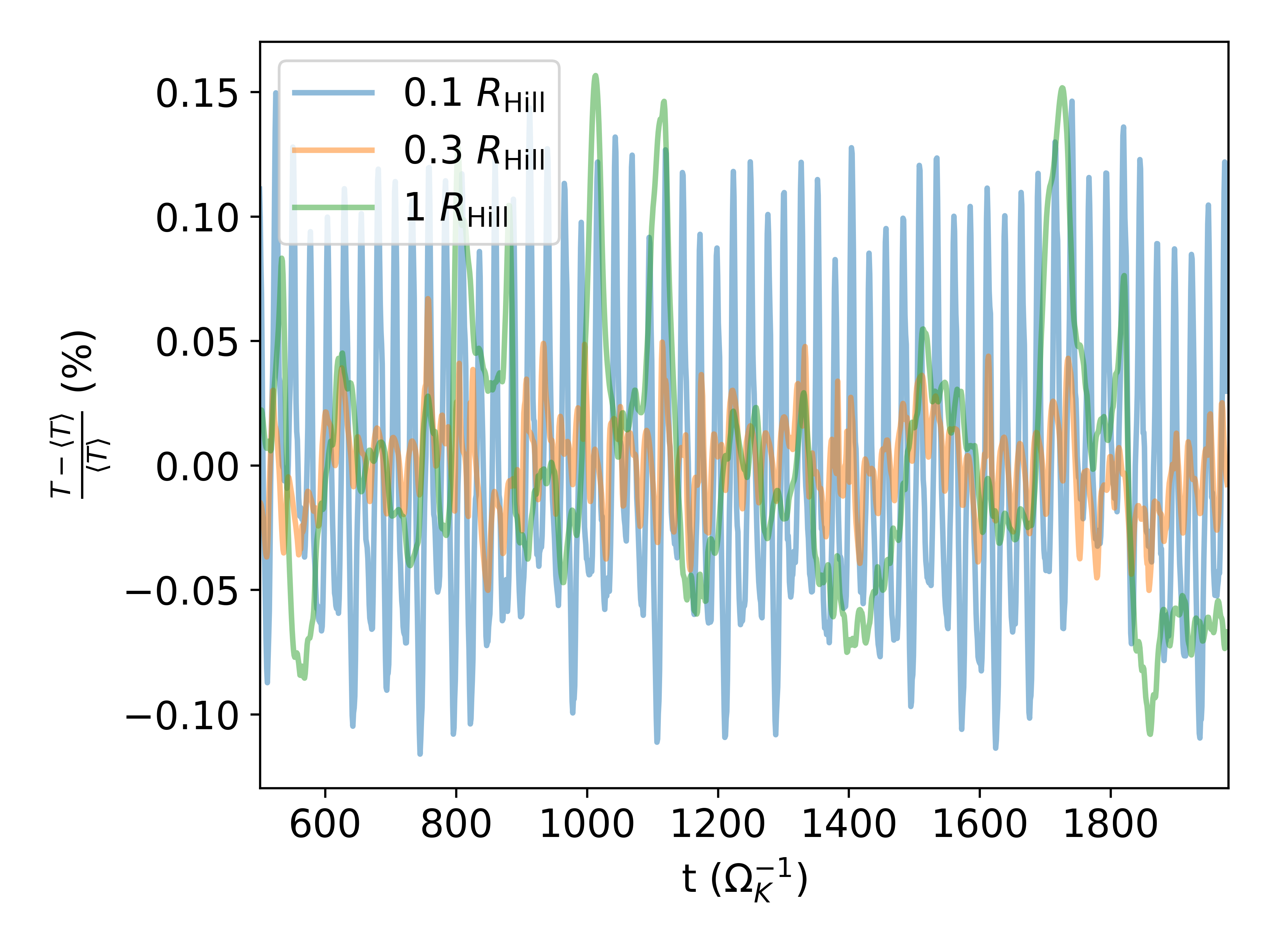}
   \caption{
      Thermal evolution of the three-dimensional simulation.
      \emph{Left:} Time evolution of the spherically averaged entropy.
      The lines are spaced logarithmically in time and the last four lines overlap, indicating the emergence of a radial entropy profile that is constant in time.
      \emph{Right:} Relative temperature deviation to the mean temperature in the steady state at three distances to the protoplanetary core. The trend line slopes of $(-0.84\pm 4.21)\cdot 10^{-6}\,\% \Omega_K$ at $0.1\,R_\mathrm{Hill}$, $(-4.21\pm 1.12)\cdot 10^{-6}\,\% \Omega_K$ at $0.3\,R_\mathrm{Hill}$, and $(-2.28\pm 0.31)\cdot 10^{-5}\,\% \Omega_K$ at $1\,R_\mathrm{Hill}$ are consistent with a true steady state.
   }
   \label{fig:entropy}%
\end{figure*}

\begin{figure*}[htbp]
   \centering
   \includegraphics[width=0.98\hsize]{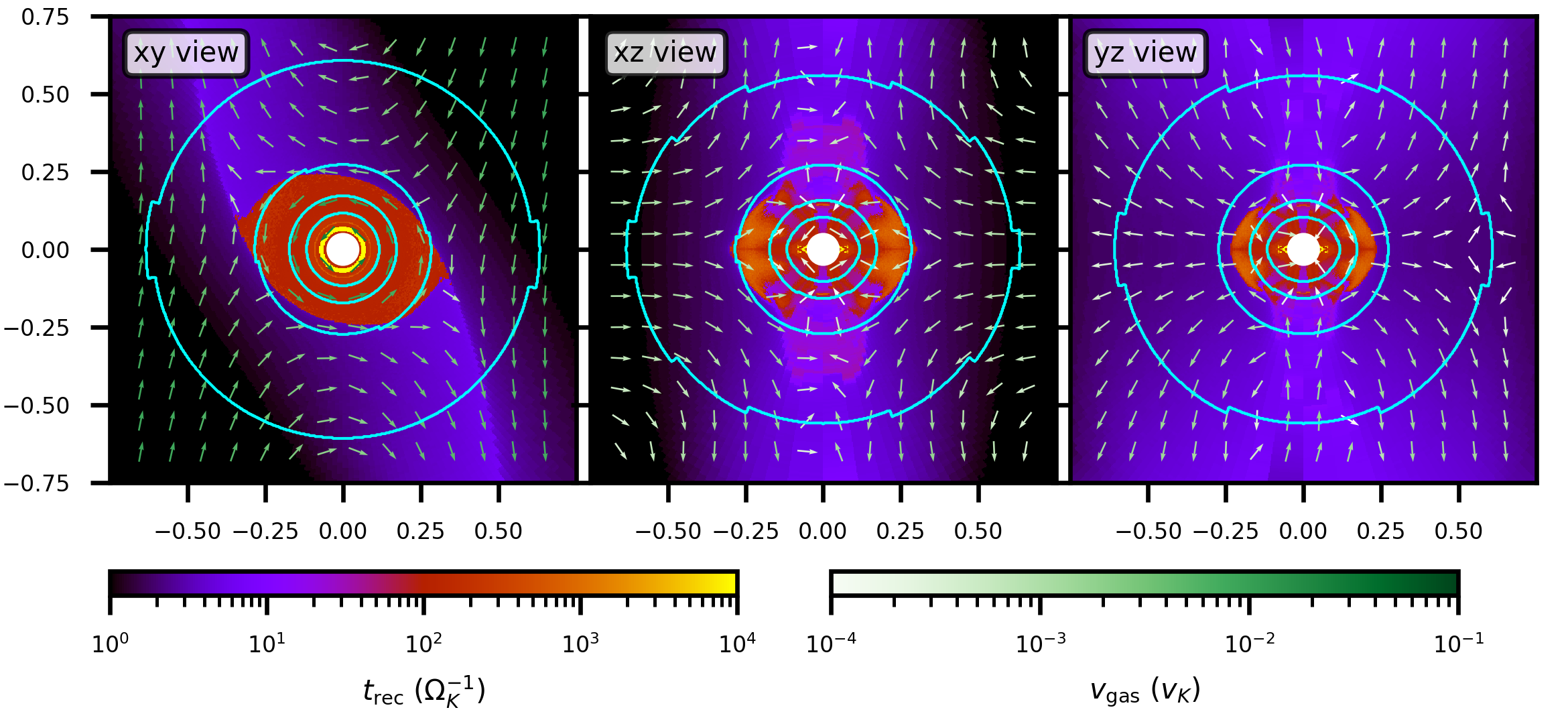}
   \caption{
      Recycling time and density of the gas inside the Hill sphere.
      The length unit is $R_\mathrm{Hill}$.
      The velocity of the gas is measured in units of the Keplerian orbital velocity of the planet around the star.
      The recycling time is defined as the time the gas has already spent inside the Hill sphere.
      The arrows indicate the velocity of the gas.
      The density contour lines (cyan) are drawn at $[2,4,8,16] \, \rho_0$.
      These lines are jagged because of the resolution of the grid.
      \emph{Left:} The xy plane or midplane.
      \emph{Middle:} The xz plane.
      \emph{Right:} The yz plane.
      The atmosphere is recycled up to the surface of the protoplanetary core.
      Gas around the $z$-axis has a lower recycling time than the gas in the midplane.
      This indicates that gas enters from the poles and leaves the atmosphere through the midplane.
   }
   \label{fig:trec}%
\end{figure*}

\begin{figure}[htbp]
   \centering
   \includegraphics[width=0.98\hsize]{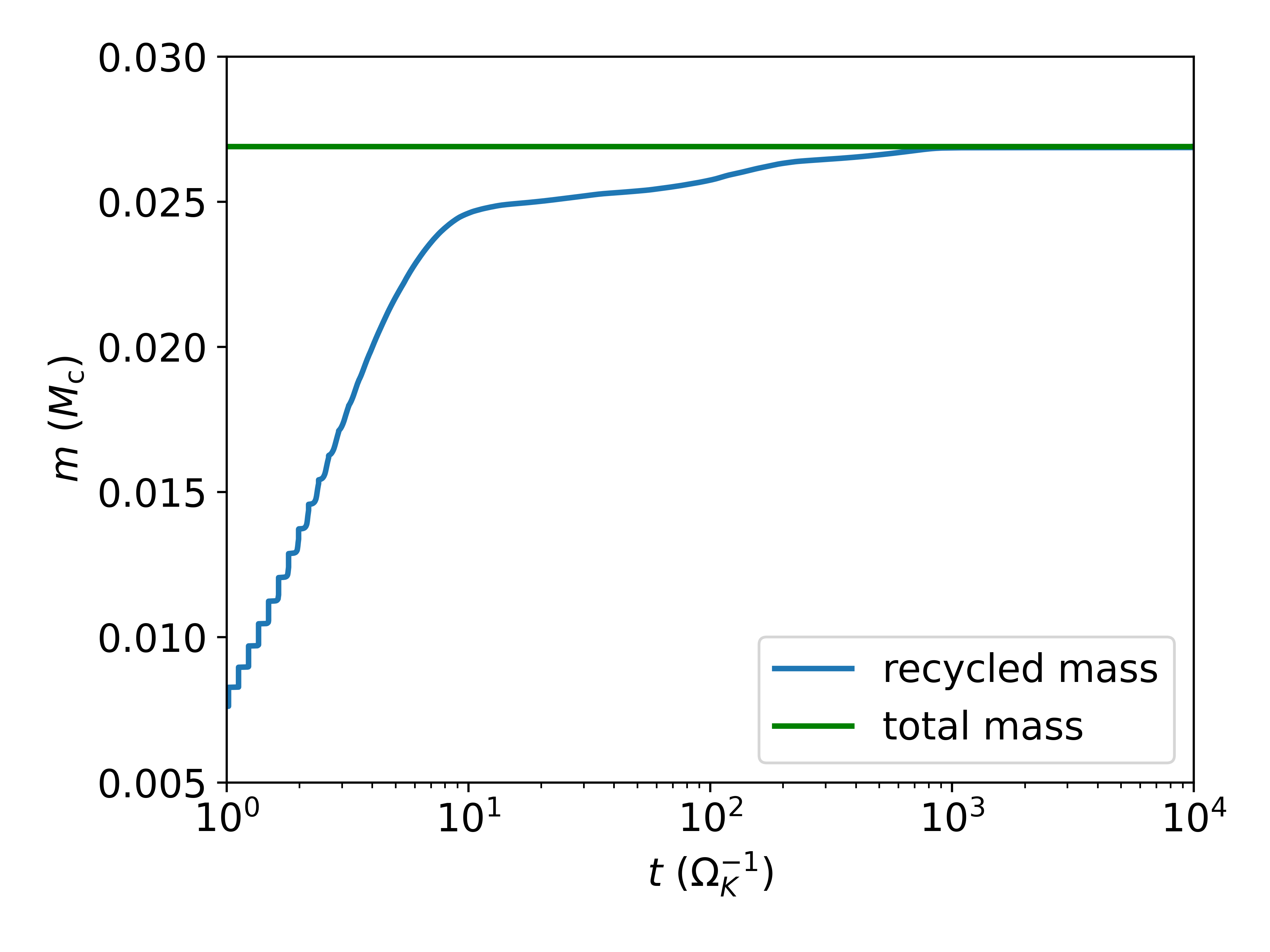}
   \caption{
      Recycled mass in the Hill sphere as a function of time after a steady state has emerged.
      Most mass in the Hill sphere is part of the shearing flow and is therefore recycled very quickly ($\approx 10\,\Omega_K^{-1}$).
      For the inner parts of the atmosphere to be recycled, it takes approximately $10^{3}\,\Omega_K^{-1}$.
      A small part of the atmosphere is not recycled even after our maximum integration time of $10^{4}\,\Omega_K^{-1}$.
   }
   \label{fig:trec_hist}%
\end{figure}

\section{Results}
We first present the comparison of the three-dimensional simulation to a one-dimensional simulation with identical physical parameters to highlight that a three-dimensional simulation behaves fundamentally differently.
In \se{result_steady_state}, we then demonstrate the steady state found in the three-dimensional simulation by analyzing the entropy evolution and show that the atmosphere eventually stops cooling.
In \se{result_recycling}, we then show,  using tracer particles, that the steady state is indeed maintained by the recycling of the atmospheric gas with gas from the circumstellar disk, in agreement with the recycling hypothesis.
\subsection{One-dimensional versus three-dimensional comparison}
\label{sec:result_comparison}

Figure\,\ref{fig:mass} displays the mass inside the Hill sphere as a function of time since the start of the simulation for the three-dimensional simulation and a one-dimensional simulation with identical parameters.
The one-dimensional simulation starts with a slightly higher mass because, unlike the three-dimensional simulation, it does not include the vertical stratification of the circumstellar disk.
Initially, during the switch-on time $t_\mathrm{inj}$, both simulations behave similarly.
However, afterward the one-dimensional simulation continues to accrete mass, while the three-dimensional simulation stops accreting.
In the three-dimensional simulation, the final ratio of the mass inside the Hill sphere and the core mass, $\chi = M_\mathrm{final}/M_c$, converges to $\chi = 2.76\,\%$, a value that is typical for a mini-Neptune planet. In the one-dimensional simulation, gas accretes on a timescale shorter than 1000 years. A mass ratio of $\chi = 100\,\%$ is reached at $t=35000\,\Omega_K^{-1}$, and $\chi=200\,\%$ is reached at $t=65000\,\Omega_K^{-1}$.
This behavior suggests that a pure three-dimensional mechanism is responsible for counteracting the radiative cooling and prevents the atmosphere from contracting.

\subsection{Steady state}
\label{sec:result_steady_state}

In Fig.\,\ref{fig:entropy}, we display the thermal evolution of the three-dimensional simulation.
The left panel shows the evolution of the radial entropy profile from the start of the simulation up until the maximum simulation time of $2000\,\Omega_K^{-1}$.
We show the entropy profile because adiabatic compression and expansion do not change the entropy of the gas, and therefore the entropy is a direct indicator for the cooling of the gas.
The entropy profile lines are spaced logarithmically in time.
Initially, the entropy drops rapidly because of the radiative cooling.
As explained in Sect. \ref{sec:model}, the adopted opacity is very low in order to facilitate a cooling timescale shorter than the simulation time.
After approximately $200\,\Omega_K^{-1}$, the entropy profile stops decreasing and becomes constant in time, ignoring small oscillations.
These oscillations are displayed in the right panel of Fig.\,\ref{fig:entropy}.
There we see three distinctive radii, $r \in [0.1, 0.3, 1] \, R_\mathrm{Hill}$, and compare the temperature to the mean temperature averaged over the steady state ($t>500\,\Omega_K^{-1}$).
With a maximum amplitude of $0.15\,\%$ in temperature and a maximum amplitude of $0.3\,\%$ in density, these oscillations are very weak and therefore not important for the evolution of the atmosphere.
Toward the inner boundary, the oscillations are strongly periodic, in contrast to the outer regions where the oscillations are more irregular.
The low frequency of the oscillations, $\sim 0.038\,\Omega_\mathrm{K}$, is not consistent with the high local Brunt-V\"ais\"ala frequency ($\sim100\,\Omega_K$), while the period is comparable to the sound crossing time through the computational domain. Hence, we conclude that they are caused by small global residual perturbations in our inviscid simulation.
The statistical trend of these oscillations is consistent with zero.
Hence, we can conclude that the three-dimensional simulation is in a true steady state in which the radiative cooling is fully compensated.

\subsection{Recycling hypothesis}
\label{sec:result_recycling}

Energy may only be transported by radiation and advection.
Radiation transports energy outward from the hotter inner regions of the atmosphere toward the cooler circumstellar disk.
Therefore, advection must counter the radiation energy flux and transport energy back inward to the planetary core.
In a steady state, where the net mass flux is zero, adiabatic compression and expansion cancel each other out.
Even though the net mass flux over a shell vanishes, this does not imply that the advection energy flux vanishes as well.
The energy of the gas that enters the atmosphere is higher than the energy of the gas that leaves the atmosphere, resulting in a net advection energy flux inward.
In one dimension, there is only one possible direction for the gas to flow, which makes recycling physically impossible.

To further investigate the recycling of the atmospheric gas, we used tracer particles to measure the time the gas has spent inside the Hill sphere, which we will refer to as the "recycling time".
The recycling time was calculated by inserting tracer particles throughout the Hill sphere.
In order to remove the small oscillations shown in Figure\,\ref{fig:entropy}, we time-averaged the velocity field in the steady state before introducing the tracer particles.
We then integrated them backward in time along streamlines and measured the time it takes for them to leave the Hill sphere.
This recycling time measures how much time the tracer particle has spent inside the Hill sphere; it quantifies how rapidly atmospheric gas and disk gas are exchanged.
Because the trajectory, and thus the measured recycling time, is highly sensitive to the starting location of the particles, we used many more particles than grid cells and took the median of adjacent particles.
This approach is justified because the particles represent a continuous fluid.
Figure\,\ref{fig:trec} displays two-dimensional cuts of this recycling time.
As expected, the recycling time is significantly higher in the inner part of the Hill sphere, where the gas rotates around the protoplanetary core, than in the outer regions where it flows past the planet.
Most importantly, however, almost the whole atmosphere recycles in a finite time frame.
Even the region directly at the surface of the planetary core is recycled.
Only a small region at the boundary of the vertical influx and the outflow in the midplane, the yellow ring in Fig.\,\ref{fig:trec}, did not recycle even after $10^{4}\,\Omega_K^{-1}$.
However, this small remaining part is negligible for the thermodynamic structure of the atmosphere.
Therefore, the gas that forms the planetary atmosphere is eventually replenished with fresh, high entropy gas from the circumstellar disk.

In addition to the recycling time, the figure also displays the density distribution using iso-density lines.
The density at the inner boundary reaches a maximum value of $\rho_\mathrm{max} = 117 \, \rho_0$.
We can see that both the long recycling time region and the high density region are squished vertically and along the orbital direction.
However, the long recycling time region is also elongated along the x-direction because of the stellar tidal forces.
Close to the vertical axis, the recycling time is also significantly shorter than in the midplane.
This shows that fresh gas enters the atmosphere through the poles and leaves the atmosphere by spiraling outward in the midplane before it gets picked up by the shearing flow.

Figure\,\ref{fig:trec_hist} shows the mass-weighted recycling time (i.e., the recycled mass) and compares it to the total mass inside the Hill sphere.
Eventually, over 99.85\,\% of the mass inside the Hill sphere is recycled with fresh gas from the circumstellar disk.
The remaining part is heated radiatively by the surrounding gas and therefore does not continue to cool either.
This illustrates that the radiative cooling is fully compensated by the recycling of the atmospheric gas with fresh gas from the circumstellar disk.

\section{Discussion and summary}

It has been customary to model planets' thermal evolution like stars, using the one-dimensional stellar structure equations. But young planets, embedded in their natal disks, cannot be regarded as isolated from their environment. Facing the disk gas flow, planets need to "process" material at a rate of $\!R_\mathrm{Hill}^3 \Omega_K \rho_0$ -- equivalent to $\sim$1 Earth mass per year (!) -- for the parameters used in this study. This number, although crude and not accounting for the the spatial variations in the recycling efficiency, in itself underscores the fact that embedded planets cannot be treated as one-dimensional isolated objects. In this letter, with improved three-dimensional RHD simulations, we have shown that the entire envelope is affected by the vigorous recycling, preventing the adiabatically heated gas from cooling radiatively. We found this despite our choice of very low opacity, for which the one-dimensional setup would enter runaway gas accretion in a mere $10^3\,\mathrm{yr}$. In a future work, we will investigate the dependence of our result on parameters such as the planet mass, location, and disk headwind. In particular, for typical disk parameters with increasing orbital radii, the impact rate of disk gas ($\sim\!R_\mathrm{Hill}^3 \Omega_K \rho_0$) decreases, while the gravitational potential at the planet surfaces steepens ($\sim\! R_\mathrm{Hill}/R$). Therefore, we can expect recycling to become less vigorous in the outer disk. Nonetheless, the ubiquity of super-Earths and mini-Neptunes close to their host star is most naturally explained by the recycling mechanism. Gas hydrodynamics trumps radiative cooling.

\begin{acknowledgements}
TWM and RK acknowledge funding through the German Research Foundation (DFG) under grant no.~KU 2849/6-1 as well as support through travel grants under the SPP 1992: Exoplanet Diversity program. 
RK acknowledges financial support via the Emmy Noether Research Group on Accretion Flows and Feedback in Realistic Models of Massive Star Formation funded by the DFG under grant no.~KU 2849/3-1 and KU 2849/3-2.
We acknowledge support by the High Performance and Cloud Computing Group at the Zentrum f\"ur Datenverarbeitung of the University of T\"ubingen, the state of Baden-W\"urttemberg through bwHPC and the DFG through grant no.~INST 37/935- 1 FUGG.
\end{acknowledgements}

\bibliography{library.bib}
\bibliographystyle{bibtex/aa}

\end{document}